\documentclass{iaus}
\usepackage{graphicx,natbib}

\newcommand{\Hipparcos}{{\sl Hipparcos}}
\newcommand{\HST}{{\sl HST}}
\newcommand{\ROSAT}{{\sl ROSAT}}
\newcommand{\Msun}{\mbox{M$_{\odot}$}}

\newcommand{\Rsun}{\mbox{R$_{\odot}$}}
\newcommand{\Mjup}{\mbox{M$_{\rm Jup}$}}

\newcommand{\hdbin}{\hbox{HD~130948BC}}
\newcommand{\hdprim}{\hbox{HD~130948A}}
\newcommand{\hdage}{\hbox{0.79$^{+0.22}_{-0.15}$~Gyr}}

\title[Confronting Substellar Theoretical Models] %% give here short title %%
{Confronting Substellar Theoretical Models with Stellar Ages}

\author[Dupuy et al.]  %% give here short author list %%
{Trent J. Dupuy$^1$, Michael C. Liu$^1$, and Michael J. Ireland$^2$}

\affiliation{$^1$Institute for Astronomy, University of Hawai`i, \\
  2680 Woodlawn Drive, Honolulu, HI 96822 USA \\ e-mail: {\tt
    tdupuy@ifa.hawaii.edu} \\[\affilskip] $^2$School of Physics,
  University of Sydney \\ NSW2006, Australia}

\pubyear{2008}
\volume{258}  %% insert here IAU Symposium No.
\pagerange{1--7}
\jname{The Ages of Stars}
\editors{E.E. Mamajek, D.R. Soderblom \& R.F.G. Wyse, eds.}
\begin{document}

\maketitle

\begin{abstract}

  By definition, brown dwarfs never reach the main-sequence, cooling
  and dimming over their entire lifetime, thus making substellar
  models challenging to test because of the strong dependence on age.
  Currently, most brown dwarfs with independently determined ages are
  companions to nearby stars, so stellar ages are at the heart of the
  effort to test substellar models. However, these models are only
  {\em fully} constrained if both the mass and age are known. We have
  used the Keck adaptive optics system to monitor the orbit of \hdbin,
  a brown dwarf binary that is a companion to the young solar analog
  \hdprim. The total dynamical mass of 0.109$\pm$0.003~\Msun\ is the
  most precise mass measurement (3\%) for any brown dwarf binary to
  date and shows that both components are substellar for any plausible
  mass ratio. The ensemble of available age indicators from the
  primary star suggests an age comparable to the Hyades, with the most
  precise age being \hdage\ based on gyrochronology. Therefore,
  \hdbin\ is unique among field L and T dwarfs as it possesses a
  well-determined mass, luminosity, and age. Our results indicate that
  substellar evolutionary models may underpredict the luminosity of
  brown dwarfs by as much as a factor of $\approx$2--3$\times$. The
  implications of such a systematic error in evolutionary models would
  be far-reaching, for example, affecting determinations of the
  initial mass function and predictions of the radii of extrasolar
  gas-giant planets. This result is largely based on the reliability
  of stellar age estimates, and the case study of \hdprim\ highlights
  the difficulties in determining the age of an arbitrary field star,
  even with the most up-to-date chromospheric activity and
  gyrochronology relations. In order to better assess the potential
  systematic errors present in substellar models, more refined age
  estimates for \hdprim\ and other stars with binary brown dwarf
  companions (e.g., $\epsilon$~Ind~Bab) are critically needed.

  \keywords{stars: brown dwarfs; techniques: high angular resolution;
    binaries: close, visual; infrared: stars}

\end{abstract}

\firstsection % if your document starts with a section,
\section{Introduction}

Theoretical models of objects below the substellar limit
($M<0.075$~\Msun) are essential for characterizing the several hundred
brown dwarfs and extrasolar gas-giant planets discovered to date.
Thus, these models have become ubiquitous in the literature, even
though empirical tests of their ability to accurately predict the
properties of brown dwarfs has been limited to only a handful of
relatively warm objects. To test substellar evolutionary models, the
input parameters of mass and age must be determined. For young brown
dwarfs, the M6.5 eclipsing binary 2MASS~J05352184$-$0546085 in the
Orion Nebula provides a unique benchmark \citep{2006Natur.440..311S}.
Prior to this year, only three binaries provided dynamical mass
measurements for field objects at or below the substellar limit: the
M8.5+M9 binary LHS~1070BC \citep{2001A&A...367..183L,
  2008A&A...484..429S}; the M8.5+M9 binary Gl~569Bab
\citep{2004astro.ph..7334O, 2006ApJ...644.1183S}; and the L0.5+L1
binary 2MASS~J0746+2000AB \citep{2004A&A...423..341B}. Recent work has
contributed several more dynamical masses for objects lower in both
temperature and mass than previously studied: the mid-L dwarf GJ~802B
\citep{gl802b-ireland}; the T5+T5.5 dwarf binary 2MASS~J1534-2952AB
\citep{2008arXiv0807.0238L}; and the L4+L4 binary \hdbin\
\citep{2008arXiv0807.2450D}. While mass measurements alone can provide
very stringent tests of theoretical models (e.g., see Liu et
al. 2008), substellar evolutionary models are only fully constrained
when both the mass and age can be determined. In fact, precise ages
are critical for such tests because brown dwarfs -- unlike stars --
never reach a main-sequence, so their properties depend very
sensitively on their age.

Of the substellar field dwarfs with measured masses, only \hdbin\ has
a precisely determined age.  These nearly-identical L dwarfs were
discovered by \citet{2002ApJ...567L.133P} as companions to the young
solar analog \hdprim\ (G2V, [Fe/H]~=~0.05).  \Hipparcos\ measured a
distance of 18.17$\pm$0.11~pc \citep{2007hnrr.book.....V} for the
primary star, which enables a very precise dynamical mass measurement
when paired with our well-determined orbital solution.

\begin{figure}[b]
  \begin{center}
  \includegraphics[height=5in, angle=90]{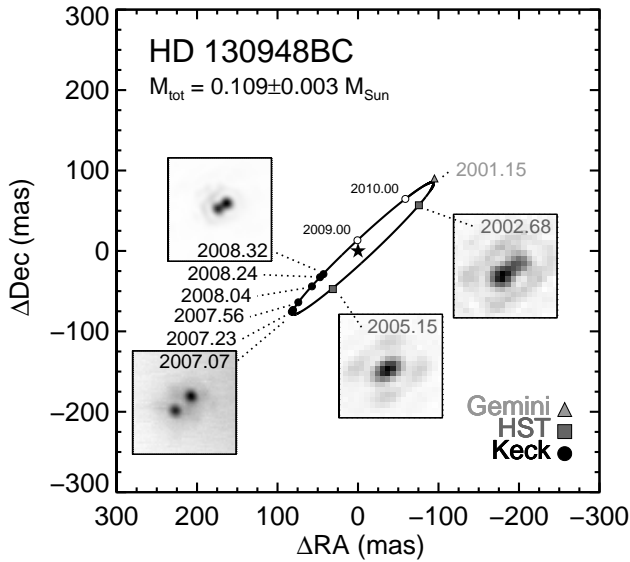}
  \caption{Keck (circles), \textit{HST} (squares), and Gemini (triangle)
    relative astrometry for \hdbin\ along with the best-fit orbit.
    Error bars are smaller than the plotting symbols. The empty
    circles are the predicted positions in 2009 and 2010.}
  \label{fig:orbit}
  \end{center}
\end{figure}

\section{The Mass of \hdbin}

We have used Keck adaptive optics (AO) imaging to monitor the relative
orbit of the two components of \hdbin\ (Figure~\ref{fig:orbit}).
Combined with archival {\sl Hubble Space Telescope} (\HST) imaging and
a re-analysis of the Gemini discovery data, our data span
$\approx$7~years ($\approx$70\% of the orbital period). We fit a
simple analytic PSF model to derive astrometry from the Keck and
Gemini images, while TinyTim model PSFs were fit to the \HST\ images.
An individually tailored Monte Carlo simulation was used to determine
the astrometric uncertainty for each observation epoch. The resulting
astrometry is extremely precise with typical Keck errors of
300~$\mu$as, corresponding to $\approx$1~\Rsun\ at the distance of
this system, while the orbit is roughly the size of the asteroid belt.
We determined the binary's orbital parameters and their confidence
limits using a Markov Chain Monte Carlo (MCMC) technique. The best-fit
orbit has a reduced $\chi^2$ of 1.06 (9~degrees-of-freedom), thus
validating our astrometric error estimates. Applying Kepler's Law to
the MCMC-derived orbital period ($P=9.9^{+0.7}_{-0.6}$~yr) and
semimajor axis ($a=121\pm6$~mas) yields a dynamical mass of
0.1089$^{+0.0020}_{-0.0017}$~\Msun. Accounting for the additional
uncertainty in the \Hipparcos\ distance results in a dynamical mass of
0.109$\pm$0.003~\Msun\ (114$\pm$3~\Mjup).

In the following analysis, we apportion the total mass between the two
components by converting the measured luminosity ratio into a mass
ratio using evolutionary models. The resulting individual masses are
very insensitive to the models used because the flux ratio is so close
to unity (the steepness of the mass--luminosity relation means that
even small differences in mass result in large differences in
luminosity). Regardless, we are careful to conduct our analysis in a
self-consistent manner free of circular logic.

\section{The Age of \hdprim}

\noindent As a young solar analog, multiple indicators are available
to assess the age of \hdprim:

\medskip
%\smallskip
\noindent $\bullet$ {{\it Rotation/Gyrochronology}} ---
\citet{2000AJ....120.1006G} measured two rotation periods of 7.69 and
7.99~days for \hdprim. Thus, we adopt a rotation period of
7.84$\pm$0.21~days and a $B-V$ color of 0.576$\pm$0.016~mag from the
\Hipparcos\ catalog. We employ the \citet{mam08-ages} calibration of
the ``gyrochronology'' relation originally introduced by
\citet{2007ApJ...669.1167B}. The age we derive is \hdage, where the
confidence limits are determined through a Monte Carlo approach in
which the period, color, and empirical coefficients are drawn from
normal distributions corresponding to their uncertainties.

\medskip
%\smallskip
\noindent $\bullet$ {{\it Chromospheric Activity}} ---
\citet{1996AJ....111..439H} and \citet{2004ApJS..152..261W} measure
$\log(R^{\prime}_{HK})$ values of $-$4.45 and $-$4.50 for \hdprim,
respectively. Using the activity--age relation of \citet{mam08-ages},
we derive ages of 0.4 and 0.6~Gyr from these $\log(R^{\prime}_{HK})$
values. The empirical relation is expected to gives ages with an
uncertainty of $\approx$0.25~dex, so we adopt a mean age of
0.5$\pm$0.3~Gyr from this method.

\medskip
%\smallskip
\noindent $\bullet$ {{\it X-ray Activity}} --- \hdprim\ was detected
by \ROSAT, and \citet{1999A&AS..135..319H} measure
$\log(L_X)$~=~29.01~dex (cgs), which gives $\log(R_X)$~=~$-$4.70.
Using the empirical relation of \citet{1998PASP..110.1259G}, this
corresponds to an age of 0.1--0.3~Gyr, depending on whether we adopt
$\alpha$ of 0.5 or $1/\exp$. The X-ray relation of \citet{mam08-ages},
derived by combining their $\log(R_X)$--$\log(R^{\prime}_{HK})$ and
$\log(R^{\prime}_{HK})$--age relations, gives an age of 0.5~Gyr. The
X-ray luminosity of \hdprim\ is in agreement with single G~stars in
the Pleiades and Hyades (28.9--29.0; \citet{1995ApJ...448..683S,
  2001A&A...377..538S}).

\medskip
%\smallskip
\noindent $\bullet$ {{\it Isochrones}} --- Using high resolution
spectroscopic data combined with a bolometric luminosity and model
isochrones, \citet{2005ApJS..159..141V} derived an age estimate of
1.8~Gyr, with a possible age range of 0.4--3.2~Gyr. From the same data
and with more detailed analysis, \citet{2006astro.ph..7235T} found a
median age of 0.72~Gyr, with a 95\% confidence range of
0.32--2.48~Gyr.

\medskip
%\smallskip
\noindent $\bullet$ {{\it Lithium}} --- Measurements by
\citet{1981ApJ...248..651D}, \citet{1985ApJ...290..284H}, and
\citet{2001A&A...371..943C} give lithium equivalent widths of
95$\pm$14, 96$\pm$3, and 103$\pm$3 m\AA, respectively, for \hdprim.
Compared to stars of similar color, these values are slightly lower
than the mean for the Pleiades and slightly higher than for UMa and
the Hyades, though consistent with the scatter in each cluster's
measurements \citet{1993AJ....106.1080S, 1993AJ....106.1059S,
  1993AJ....105.2299S}

\medskip
%\smallskip 

In summary, the most precise age estimate available for \hdprim\ comes
from gyrochronology, which gives an age of \hdage. All other age
indicators agree with this estimate, though this is due to their large
uncertainties rather than a true consensus.

\section{Substellar Evolutionary Models Fully Constrained}

\begin{figure}
  \begin{center}
  \includegraphics[height=.6\textheight, angle=90]{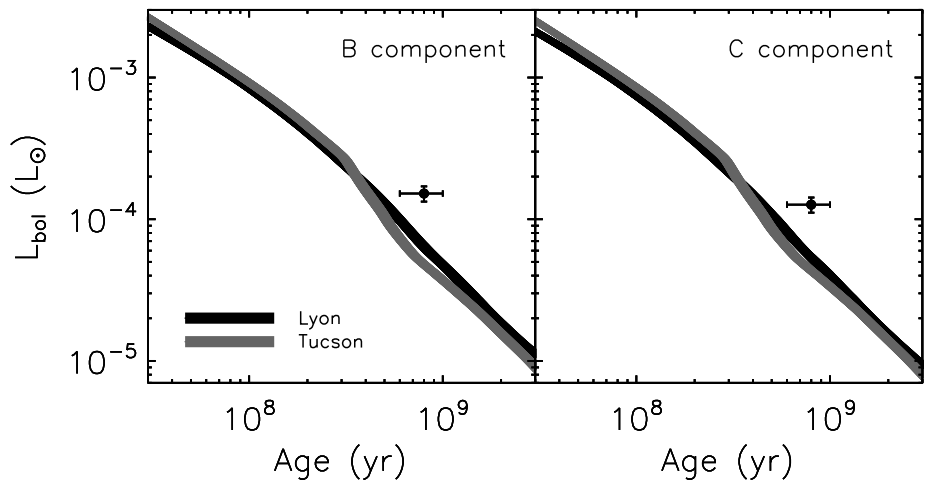}
  \caption{The filled circles mark the measured luminosities of
    HD~130948B and C at the age we derive for HD~130948A. The thick
    shaded lines are isomass lines from evolutionary models, where the
    line thicknesses encompass the 1$\sigma$ errors in the individual
    masses of \hdbin. Although the two independent sets of models
    agree very well with one another, they underpredict the
    luminosities of \hdbin\ by a factor of $\approx$2--3$\times$.}
  \label{fig:lbol-age}
  \end{center}
\end{figure}

With a measured mass, luminosity, and age, \hdbin\ provides the first
direct test of the luminosity evolution predicted by theoretical
models for substellar field dwarfs. Both the Tucson models
\citep{1997ApJ...491..856B} and Lyon models
\citep[DUSTY;][]{2000ApJ...542..464C} underpredict the luminosities of
HD~130948B and C given their masses and age. The discrepancy is quite
large, about a factor of 2 for the Lyon models and a factor of 3 for
the Tucson models (Figure~\ref{fig:lbol-age}). If the age and
luminosities of HD~130948B and C had been used to infer their masses,
the resulting estimates would have been too large by 20--30\%. In
order to explain this discrepancy entirely, model radii would have to
be underpredicted by 30--40\%. Alternatively, the age of \hdprim\
would need to be $\approx$0.4~Gyr in order to resolve this
discrepancy. Although such a young age is marginally consistent with
the various age indicators; it is on the extreme young end of two
independent, well-calibrated age estimates (gyrochronology and stellar
isochrones). In order to better assess this discrepancy between models
and data, a more refined age estimate for \hdprim\ (e.g., from
asteroseismology) is critically needed.

\section{Lithium Depletion in \hdbin}

\begin{figure}
  \begin{center}
  \includegraphics[height=.5\textheight, angle=0]{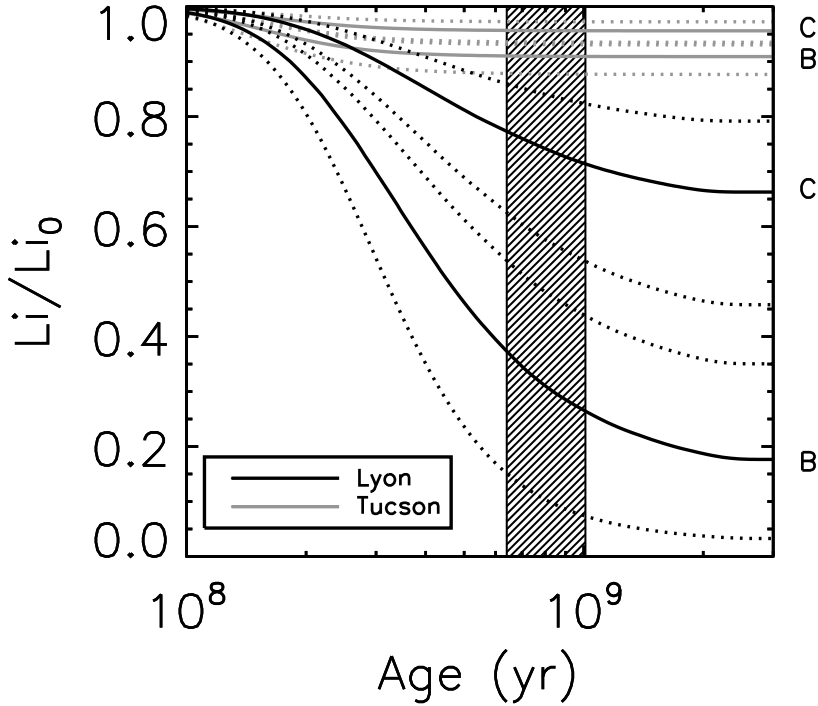}
  \caption{Lithium depletion as a function of age as predicted by
    evolutionary models.  The solid lines correspond to the individual
    masses of HD~130948B and C.  These lines are bracketed by dotted
    lines that correspond to the 1$\sigma$ uncertainties in the
    individual masses.  The ordinate is the fraction of initial
    lithium remaining.  The hatched black box indicates the constraint
    from the age of \hdprim\ estimated from gyrochronology.}
  \label{fig:lithium}
  \end{center}
\end{figure}

Since brown dwarfs are fully convective objects, they can rapidly
deplete their initial lithium if their core temperature is ever high
enough to do so. This threshhold is reached around 0.065~\Msun, and
since this is below the hydrogen-burning mass-limit, this fact has
been exploited to identify sufficiently old objects bearing lithium as
substellar. In fact, the exact mass-limit for lithium burning is
slightly different depending on which sets of theoretical models are
used, and the masses of HD~130948B and C happen to be very close to
these theoretically predicted mass-limits
(Figure~\ref{fig:lithium}). According to the Tucson models, neither
component is massive enough to have ever depleted a significant amount
of its initial lithium. The Lyon models, on the other hand, predict
that HD~130948B is massive enough to have depleted most of its
lithium, while HD~130948C is not. Thus, resolved optical spectroscopy
designed to detect the lithium doublet at 6708~\AA\ would provide a
very discriminating test of substellar evolutionary models, which are
otherwise nearly indistinguishable (e.g., see
Figure~\ref{fig:lbol-age}). This experiment can currently only be
conducted with \HST/STIS given the very small binary separation
($<130$~mas).

\section{Future Prospects}

Brown dwarfs hold the potential to address many astrophysical
problems. For example, they are excellent laboratories in which to
study ultracool atmospheres under a variety of conditions, and they
may eventually be useful as Galactic chronometers given how
sensitively their properties depend on their age (see contribution by
A. Burgasser). However, the theoretical models we rely upon to
characterize brown dwarfs have only begun to be rigorously tested by
benchmark systems such as \hdbin.  More results are expected to be
forthcoming over the next several years for other brown dwarf binaries
with stellar companions: $\epsilon$~Ind~Bab
\citep{2004A&A...413.1029M}; Gl~417BC \citep{2003AJ....126.1526B};
and GJ~1001BC \citep{2004AJ....128.1733G}. However, the utility of
these systems as benchmarks critically depends on the confidence in
the age estimates for their primary stars. Therefore, these stars
deserve special attention so that state-of-the-art age-dating
techniques (e.g., asteroseismology and gyrochronology) may be applied
to them. Also, extending the empirical relations between age, stellar
rotation, and chromospheric activity to include objects with as late a
spectral type as possible will enable many more systems to be used as
benchmarks for testing models.  These relations are currently only
calibrated for stars as late as early-K, but about half of the stars
with brown dwarf companions have spectral types between early-K and
early-M.

%\clearpage

\newcommand\aj{\textit{AJ}}%         % Astronomical Journal 
\newcommand\araa{\textit{ARAA}}%         % Annual Review of Astron and Astrophys 
\newcommand\apj{\textit{ApJ}}%         % Astrophysical Journal 
\newcommand\apjl{\textit{ApJ} (Letters)}%         % Astrophysical Journal, Letters 
\newcommand\apjs{\textit{ApJS}}%         % Astrophysical Journal, Supplement 
\newcommand\aap{\textit{A\&A}}%         % Astronomy and Astrophysics 
\newcommand\aapr{\textit{A\&AR}}%         % Astronomy and Astrophysics Reviews 
\newcommand\aaps{\textit{A\&AS}}%         % Astronomy and Astrophysics, Supplement 
\newcommand\baas{\textit{BAAS}}%         % Bulletin of the AAS 
\newcommand\mnras{\textit{MNRAS}}%         % Monthly Notices of the RAS 
\newcommand\nat{\textit{Nature}}%         % Nature 
\newcommand\pasp{\textit{PASP}}%         % Publications of the ASP 

%\bibliography{/Users/tdupuy/tex/bibtex/tdupuy}
%\bibliographystyle{apj}

\begin{discussion}

  \bigskip
  \noindent {\bf F. Walter:} It is always risky to attempt to pin down
  the age of a field star, even using multiple techniques that may
  agree.  How much would the age have to be changed to place the
  L~dwarfs on the proper evolutionary tracks?

  \smallskip
  \noindent {\bf T. Dupuy:} I agree and would really like to see
  another independent measurement of the age, for example, from
  asteroseismology.  The age of the system would have to be about
  0.4~Gyr to bring the models into agreement with the data.

%%%%%%%%%%%%%%%%%%%%%%%%

  \bigskip
  \noindent {\bf E. Jensen:} Is there a measured metallicity for
  HD~130948A?  The metallicity will affect the evolutionary models,
  both in the H-R diagram and the predicted Li depletion.

  \smallskip
  \noindent {\bf T. Dupuy:} That's exactly right; a detail I didn't go
  into.  The metallicity of HD~130948A is basically solar, which means
  we can use the standard models.  This is another reason why having
  brown dwarfs with stellar companions is great -- because you can
  make sure you're not being confused by metallicity effects like
  those Adam talked about.

%%%%%%%%%%%%%%%%%%%%%%%%

  \bigskip
  \noindent {\bf A. West:} Does the fact that this system is a close
  binary affect the measured luminosity (because the radii are
  affected)?

  \smallskip
  \noindent {\bf T. Dupuy:} The binary separation is about 2.2 AU, so
  it's unlikely that tidal effects are at work in this system.  Also,
  it turns out that the two components receive about as much flux from
  each other as they do from the primary star, so irradiation
  shouldn't be affecting them much.

%%%%%%%%%%%%%%%%%%%%%%%%%

  \bigskip
  \noindent {\bf J. Fernandez:} The next main source of benchmarks for
  brown dwarfs will be Kepler and Corot.  The precise determination of
  ages for these primary stars will be crucial.

\end{discussion}

\end{document}